# Do L chondrites come from the Gefion family?

Allison M. McGraw,[1]⋆ Vishnu Reddy[1] and Juan A. Sanchez[2]

[1]*Lunar and Planetary Lab, The University of Arizona, 1629 E University Blvd, Tucson, AZ 85721, USA*
[2]*Planetary Science Institute, 1700 E Ft. Lowell, Tucson, AZ 85719, USA*

**ABSTRACT**
Ordinary chondrites (H, L, and LL chondrites) are the most common type of meteorites comprising 80 per cent of the meteorites that fall on Earth. The source region of these meteorites in the main asteroid belt has been a basis of considerable debate in the small bodies community. L chondrites have been proposed to come from the Gefion asteroid family, based on dynamical models. We present results from our observational campaign to verify a link between the Gefion asteroid family and L chondrite meteorites. Near-infrared spectra of Gefion family asteroids (1839) Ragazza, (2373) Immo, (2386) Nikonov, (2521) Heidi, and (3860) Plovdiv were obtained at the NASA Infrared Telescope Facility (IRTF). Spectral band parameters including band centres and the band area ratio were measured from each spectrum and used to constrain the composition of these asteroids. Based on our results, we found that some members of the Gefion family have surface composition similar to that of H chondrites, primitive achondrites, and basaltic achondrites. No evidence was found for L chondrites among the Gefion family members in our small sample study. The diversity of compositional types observed in the Gefion asteroid family suggests that the original parent body might be partially differentiated or that the three asteroids with non-ordinary chondrite compositions might be interlopers.

**Key words:** methods: observational – techniques: spectroscopic.

## 1 INTRODUCTION

Linking meteorites to their source asteroids is an important goal of planetary sciences. At present, there are less than 10 meteorite types that have been linked to their parent bodies (e.g. Gaffey et al. 1993; Vernazza et al. 2014; Reddy et al. 2015). For example, lunar samples collected from the Apollo missions have been linked to lunar meteorites, Martian meteorites have been linked to surface compositions through information gathered from Martian landers, and HED meteorites (howardites, eucrites, and diogenites) have been compositionally linked to the asteroid (4) Vesta (e.g. Reddy et al. 2012). Ground-based spectral characterization of asteroids would enable us to link these objects with their meteorite analogues in our collection (e.g. Reddy et al. 2012; Sanchez et al. 2015).

Ordinary chondrites (OC) are the dominant meteorite type to fall on Earth and are likely derived from some S-type asteroids. H, L, and LL chondrites are the three types of OC and differ in their metal abundance, silicate composition, and reduction-oxidation (redox) state. A number of parent bodies/families have been proposed for OC (Vernazza et al. 2014). These include asteroid (6) Hebe or the Koronis family for the H chondrites (Gaffey & Gilbert 1998;

⋆ E-mail: ammcgraw@email.arizona.edu

Sanchez et al. 2015), the Gefion asteroid family for L chondrites (Nesvorny et al. 2009), and the Flora family for LL chondrites (Vernazza et al. 2009). Identifying the formation location of the OC is important for understanding the thermal and redox gradients of the early Solar system.

Constraining the distribution of thermal and redox gradients across the inner Solar system would help us understand the conditions under which our terrestrial planets formed. Silicate mineralogy assemblages often depict planetary processes; they leave behind information that represents temperature and pressure conditions inside a small body or a planet at a given time in its history. By determining the source regions for silicate meteorites such as H, L, LL, primitive achondrites, and basaltic achondrites (BA), a more reasoned understanding of the mineralogical gradient of the Solar system may be obtained. Each of the silicate meteorite types depicts a unique, yet significant, history that indicates different geological associations. For instance, the significance of BA is important when considering that basalt covers 80 per cent of Earth's surface. Some primitive achondrite meteorites show partial melting features which require high temperature and pressure environments to induce melting of the pyroxene minerals present, likely within a parent body.

The Gefion family has been proposed to be the source of the L chondrite meteorites by various authors (Nesvorny et al. 2009; Vernazza et al. 2014). The family lies close to the 5:2 mean motion resonance with Jupiter, given the semimajor axis range of 2.7–2.82 au (Nesvorny et al. 2009). Fig. 1 shows the proper eccentricity and proper inclination versus semimajor axis for asteroids belonging to this family along with the objects studied in this work. Orbital elements for the five observed asteroids in this campaign compared with asteroid (1272) Gefion are shown in Table 1. While a dynamic asteroid family shares proper orbital elements, a genetic asteroid family shares both orbital proper elements as well as compositional characteristics. The Gefion family has been proposed to share both of these parameters (Nesvorny et al. 2009). This family may also be responsible for fossil L chondrite meteorites found in Ordovician Limestone (Haack et al. 1996). Haack et al. (1996) argued that a catastrophic collision 500 million year ago ejected L chondrite material from the parent body into an Earth crossing orbit causing a greater flux of these meteorites at that time. The estimated age of the fossil meteorites is roughly 467 million years, and would have impacted near a critical time for biologically and evolutionary changes for life on the Earth.

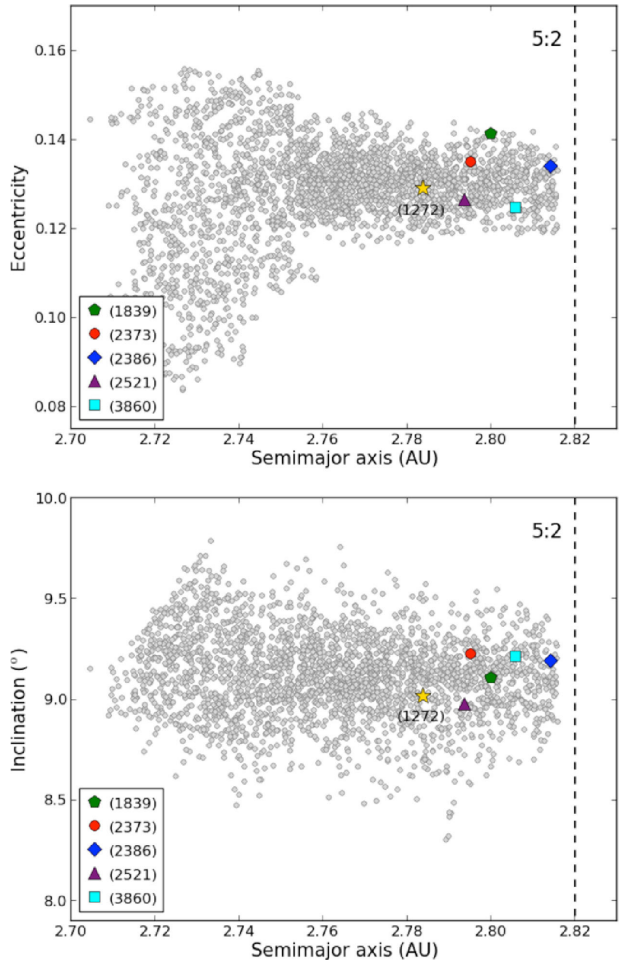

**Figure 1.** Proper eccentricity and proper inclination versus semimajor axis for (1839) Ragazza, (2373) Immo, (2386) Nikonov, (2521) Heidi, and (3860) Plovdiv. Asteroids belonging to the Gefion family from Nesvorny (2012) are depicted as grey circles. As a reference, asteroid (1272) Gefion is also included and depicted with a star. The location of the 5:2 mean motion resonance is represented by a vertical dashed line.

**Table 1.** Orbital elements semimajor axis (au), eccentricity, and inclination (degrees) for the five observed asteroids in this observational campaign shown with asteroid (1272) Gefion.

| Asteroid | Semimajor axis (au) | Eccentricity | Inclination (degrees) |
| --- | --- | --- | --- |
| (1272) Gefion | 2.7840 | 0.1521 | 8.4185 |
| (1839) Ragazza | 2.7989 | 0.1680 | 10.1663 |
| (2373) Immo | 2.7958 | 0.1715 | 10.0789 |
| (2386) Nikonov | 2.8148 | 0.1583 | 9.0851 |
| (2521) Heidi | 2.7917 | 0.0938 | 7.7235 |
| (3860) Plovdiv | 2.8069 | 0.1548 | 8.1208 |

In this paper, we present results of our observational campaign to verify a link between the Gefion asteroid family and L chondrite meteorites as proposed by Nesvorny et al. (2009) and Vernazza et al. (2014). Our goal was to use the mineralogy of the observed asteroids derived from their near-infrared (NIR) spectra to verify the link with L chondrite meteorites. We used laboratory spectral calibrations of OC to confirm our results. These new data increase the number of Gefion family members studied so far, which allow us to discuss our results in the context of those presented in previous work (Blagan 2012; Roberts et al. 2013, 2014, 2015, 2017).

## 2 OBSERVATIONS AND DATA REDUCTION

### 2.1 Observations

We observed five asteroids belonging to the Gefion family as defined by Carruba et al. (2003) using the NASA IRTF on Mauna Kea, Hawai'i between 2014 September and 2015 September. NIR spectra (0.7–2.5 µm) of asteroids (1839) Ragazza, (2373) Immo (2386) Nikonov, (2521) Heidi, and (3860) Plovdiv were obtained using the SpeX instrument (Rayner et al. 2003). Observational circumstances for each of the asteroids are listed in Table 2. Spectra were collected using the nodding technique, where the asteroid target is switched between two different slit positions, A and B. Standard stars and solar analogue stars were also observed in order to correct for telluric features and correcting the spectra for solar continuum. The asteroid targets have an exposure time of 200 s.

### 2.2 Data reduction

The data reduction process utilized the IDL-based software package known as SPEXTOOL (Cushing et al. 2004). SPEXTOOL is a data reduction package provided by the NASA IRTF for reducing SpeX data. The data reduction protocols include: removal of the background sky by subtracting the A-B spectral pairs, flat-field calibration, cosmic ray removal, wavelength calibration, dividing the asteroid spectra by spectra of the respective solar analogue star observation (Table 2), and finally median combining individual spectra to get the final spectrum.

The initial steps of the data reduction process within SPEXTOOL involved creating master flat images, extracting the 2D spectra from the image, and combining the individual spectra from a set of observations. The next step is telluric correction where the asteroid spectrum is ratioed to a local standard star spectrum. During telluric correction, subpixel offset is also carried out to account for instrumental flexure. The solar analogue star is treated the same way as the asteroid and the resulting ratio between the solar analogue star and the local standard star is used to derive the solar continuum correction by fitting a low-order polynomial. The final step is dividing the asteroid spectrum with the solar continuum curve. The average spectra of our five asteroids are shown in Fig. 2 and are offset vertically for clarity.

**Table 2.** Observational circumstances for the asteroids observed for this campaign.

| Target | Observation date UTC (GMT) | Solar analogue star | Airmass | α (degrees) | R (au) | Mag. (V) | No. of Spectra |
| --- | --- | --- | --- | --- | --- | --- | --- |
| (1839) Ragazza | 2014/11/24 17:00:00 | SAO 93936 | 1.031 | 21.6 | 2.551 | 17.0 | 12 |
| (2373) Immo | 2014/09/23 23:00:00 | SAO 93936 | 1.345 | 23.6 | 2.505 | 17.1 | 32 |
| (2386) Nikonov | 2015/01/17 23:00:00 | SAO 120107 | 1.087 | 17.0 | 3.162 | 17.3 | 6 |
| (2521) Heidi | 2014/09/04 10:00:00 | SAO 75021 | 1.011 | 15.6 | 3.013 | 16.7 | 10 |
| (3860) Plovdiv | 2015/09/22 17:00:00 | SAO 93936 | 1.322 | 22.8 | 2.384 | 15.8 | 14 |

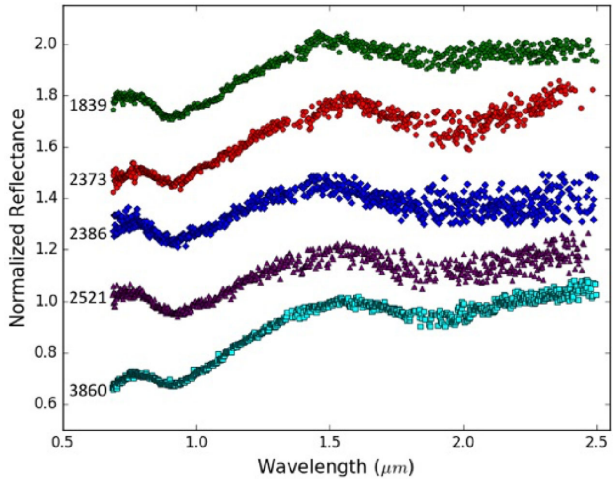

**Figure 2.** NIR (0.7–2.5 µm) spectral data of asteroid targets (1839) Ragazza, (2373) Immo, (2386) Nikonov, (2521) Heidi, and (3860) Plovdiv. The observations were obtained from the NASA IRTF using the SpeX instrument.

**Table 3.** The measured band parameters values for the Band I centre, Band II Center, and the BAR.

| Target | Band I centre (µm) | Band II centre (µm) | BAR |
| --- | --- | --- | --- |
| (1839) Ragazza | 0.93(+/−)0.01 | 1.86(+/−)0.01 | 0.76(+/−)0.15 |
| (2373) Immo | 0.93(+/−)0.01 | 1.97(+/−)0.01 | 1.56(+/−)0.15 |
| (2386) Nikonov | 0.92(+/−)0.01 | 1.99(+/−)0.01 | 1.24(+/−)0.15 |
| (2521) Heidi | 0.95(+/−)0.01 | 1.91(+/−)0.01 | 0.98(+/−)0.15 |
| (3860) Plovdiv | 0.93(+/−)0.01 | 1.91(+/−)0.01 | 1.31(+/−)0.15 |

## 3 RESULTS

All spectra shown in Fig. 2 exhibit absorption features centred at roughly 0.9 µm (Band I) and 1.9 µm (Band II). These features are attributed to crystal field transitions in $Fe^{2+}$ in minerals olivine and pyroxene. From each spectrum, band centres and the band area ratio (BAR) were measured using a PYTHON code following the protocols of Cloutis et al. (1986). Band centres correspond to the position of the minimum value in reflectance of a continuum removed absorption feature, and are determined by fitting third- and fourth-order polynomials over the bottom of each band. The BAR is given by the ratio of the area of Band II versus Band I. The band parameters were measured multiple times and from these measurements the mean values and $1\sigma$ errors were calculated. The band parameters and their uncertainties are presented in Table 3.

We plotted the Band I centre and BAR of our target asteroids in the S-asteroid subtype plot from Gaffey et al. (1993), as shown in Fig. 3. The distribution of these parameters across many spectral meteorite types suggests a wide range of silicate compositions and abundances. Asteroids (1839) Ragazza and (2521) Heidi have spectral properties and features that are similar to S(IV) subtype (analogous to OC), and in particular they fall into the H chondrite zone. Asteroids (2386) Nikonov and (3860) Plovdiv plot in the S(VI) zone. These objects are spectrally similar to silicate inclusions in iron meteorites or residue left behind after partial melting (Gaffey et al. 1993). Asteroid (2373) Immo plots in the BA region with spectral properties similar to asteroid (4) Vesta.

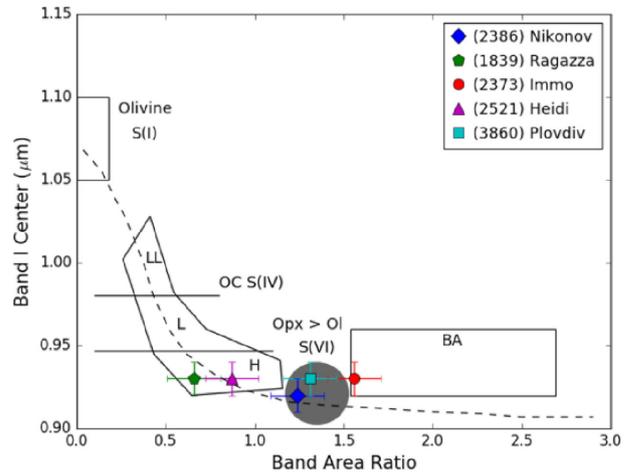

**Figure 3.** Band I centre versus BAR for the five asteroids observed as part of this survey. The polygonal region corresponds to the S(IV) subgroup defined by Gaffey et al. (1993). The rectangular zone (BA) includes the pyroxene-dominated BA assemblages. The dashed line indicates the location of the olivine-orthopyroxene mixing line of Cloutis et al. (1986). Asteroids (1839) Ragazza and (2521) Heidi plot in the S(IV) region, (3860) Plovdiv and (2386) Nikonov in the S(VI) region, and (2373) Immo in the BA region.

## 4 ANALYSIS

Based on our small sample study, the mafic mineral composition of these Gefion family asteroids suggests most are not analogous to L chondrites. Instead, they belong to a range of compositions from H chondrites, primitive achondrites, and BA.

We found that the olivine and pyroxene chemistry are consistent with that of H chondrites for asteroids (1839) Ragazza and (2521) Heidi. H chondrites are OC that possess higher iron content, roughly 25 per cent molar content, and have higher pyroxene ratios than olivine content, as they are thought to have formed under reduced conditions compared to L and LL chondrites. We used laboratory spectral calibrations derived by Dunn et al. (2010) to determine the surface composition of (1839) Ragazza and (2521) Heidi. These calibrations are valid only for asteroids that plot in the S(IV) region of the plot. Using this calibration technique, we derived their olivine and pyroxene chemistry, which are given by the molar content of fayalite (Fa) and ferrosilite (Fs), respectively. Their Fa and Fs values are consistent with H chondrites rather than L chondrites (Fig. 4). Similarly, the olivine abundances defined as the fraction olivine content to the olivine plus pyroxene content (ol/(ol+px)) for these two asteroids were found to be consistent with H chondrites, as shown in Fig. 5.

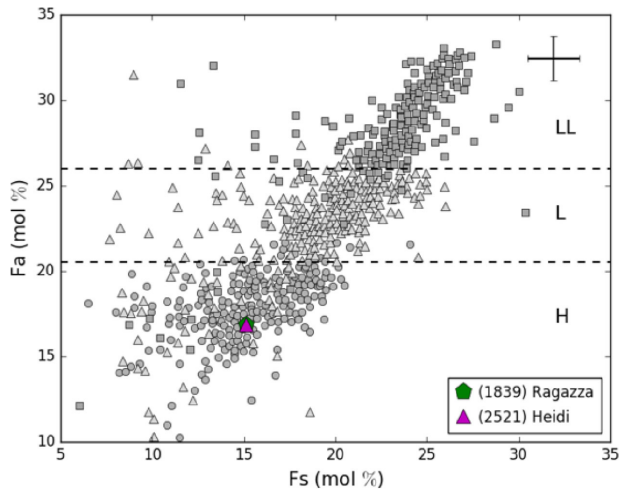

**Figure 4.** Molar contents of fayalite (Fa) versus ferrosilite (Fs) for Ragazza (green pentagon) and Heidi (magenta triangle), along with the measured values for LL (squares), L (triangles), and H (circles) OC from Nakamura et al. (2011). This plot indicates that Heidi and Ragazza are consistent with molar content of fayalite versus ferrosilite that is measured in H chondrite meteorites. The error bars in the upper right corner correspond to the uncertainties derived by Dunn et al. (2010), 1.3 molar percent for Fa, and 1.4 molar percent for Fs.

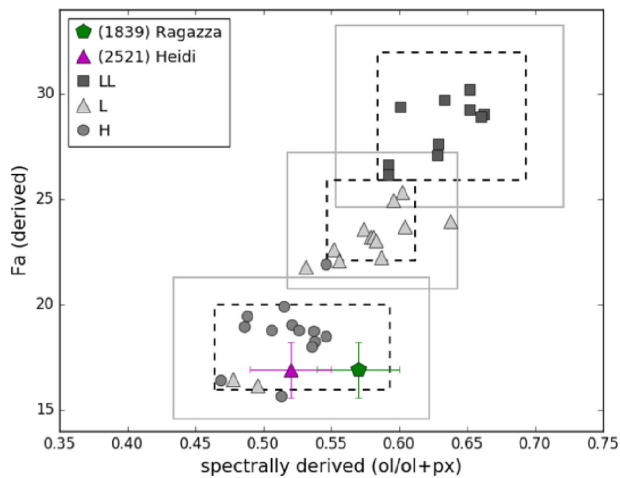

**Figure 5.** Molar content of fayalite versus ol/(ol+px) for (1839) Ragazza (green pentagon) and (2521) Heidi (magenta triangle), along with the measured values for LL, L, and H OC from Dunn et al. (2010). The molar contents of Fa versus ol/(ol+px) of asteroids Heidi and Ragazza are consistent with H chondrite meteorites.

The mineralogy of asteroids (2386) Nikonov and (2860) Plovdiv is suggestive of surfaces that are high in orthopyroxene content, and considerably low in olivine content. Given this pyroxene-dominated assembly, (2386) Nikonov and (2860) Plovdiv are consistent with primitive achondrites. Possible meteorite analogues for Nikonov and Plovdiv would likely include primitive achondrites such as acapulcoites or the slightly coarser grained lodranites, which have likely experienced more partial melting located deeper within their respective parent body (Gaffey et al. 1993). Furthermore, previous work done by Lucas et al. (2016) with the Hungaria asteroid family has found evidence for background asteroids with compositions that are consistent with primitive achondrites. It should not be ruled out that asteroids (2386) Nikonov and (2860) Plovdiv could be background asteroids to that of the Gefion family, belonging to a different asteroid family altogether.

Mineralogical analysis of asteroid (2373) Immo reveals a surface consistent with high abundances of orthopyroxene similar to that of BA. While HED meteorites derived from Vesta are the dominant achondrites in the terrestrial collection, ground-based observations (e.g. Hardersen et al. 2015) have shown non-Vesta basaltic families to exist in the main asteroid belt. Hardersen et al. (2015) propose that some Vesta-type asteroids from the inner location of the main asteroid belt could be from non-Vesta basaltic parent bodies, but could have also possibly derived from Vesta. It is possible that the asteroid (2372) Immo was derived from one such uncertain asteroid family, and is perhaps unrelated to Vesta.

We have not found any evidence for L chondrites in the Gefion family based on our small sample study. Vernazza et al. (2014) have previously suggested that the Gefion family is a mixture of both H chondrites and L chondrites. The observations in this study partly support this finding as asteroids (1839) Ragazza and (2521) Heidi have compositions that are consistent with that of H chondrite meteorites. These results also support previous work done by Roberts et al. (2013, 2014, 2015, 2017) and Blagen & Gaffey (2012). These authors have suggested that Gefion asteroid family asteroids display a varied range of compositions ranging from H, L, and LL chondrites. Based on the range of the observed compositions, we can suggest that the parent body of the Gefion family was a partially differentiated asteroid with a mixture of unaltered chondrite material, partial melt residues, and BA. Alternatively, the non-OC asteroids in the Gefion family could be interlopers that are unrelated to the original family.

## 5 CONCLUSIONS

Finding asteroid-meteorite linkage is vital for improving our understanding of the conditions in the early Solar system. In this study, we attempted to test one such link between the Gefion asteroid family and the L chondrite meteorites proposed by dynamical studies (Nesvorny et al. 2009). Based on our small sample study here is what we can summarize:

(i) the five asteroids we observed, (1839) Ragazza, (2373) Immo, (2386) Nikonov, (2521) Heidi, and (3860) Plovdiv, all have spectral properties unlike L chondrites. These asteroids have properties similar to H chondrites, primitive, and BA.

(ii) The diversity of compositional types observed in the Gefion asteroid family suggests that the original parent body might be partially differentiated.

(iii) Alternatively, the family might be contaminated with large number of interlopers from other families creating a diverse set of compositions.

(iv) Our results are consistent with similar studies of the Gefion asteroid family by Blagan et al. (2012) and Roberts et al. (2013, 2014, 2015, 2017).

## ACKNOWLEDGEMENTS

We thank the IRTF TAC for awarding time to this project, and the IRTF TOs and MKSS staff for their support. The IRTF is operated by the University of Hawaii under contract no. NNH14CK55B with the National Aeronautics and Space Administration. This work is funded by the NASA Planetary Geology and Geophysics Grants NNX14AN05G (PI: Gaffey); NNX14AM94G (PI: Reddy); NASA

Planetary Astronomy Grant NNX14AJ37G (PI: Hardersen) and supported by the NASA Undergraduate Space Grant Consortium. The authors would like to thank the reviewer for the suggestions and edits that have helped improve this manuscript.

This paper has been typeset from a TeX/LaTeX file prepared by the author.